\documentclass[twoside,fleqn]{article}
\usepackage{espcrc2,amssymb,hyperref}
\usepackage[dvips]{graphicx}

\title{Gasser--Leutwyler coefficients: A progress report}
 
\author{
  George T.~Fleming\address{
    Physics Department,
    The Ohio State University,
     Columbus, OH 43210-1168, USA.
  }, Daniel R.~Nelson$^\mathrm{a}$ and Gregory W.~Kilcup$^\mathrm{a}$
}

\begin{document}

\pagestyle{empty}

\begin{abstract}

Last year, we reported our first results on the determination
of Gasser--Leutwyler coefficients using partially quenched lattice QCD
with three flavors of dynamical staggered quarks.  We give an update
on our progress in determining two of these coefficients, including
an exhaustive effort to estimate all sources of systematic error.
At this conference, we have heard about algorithmic techniques
to reduce staggered flavor symmetry breaking and a method
to incorporate staggered flavor breaking into the partially quenched
chiral Lagrangian.  We comment on our plans to integrate these developments
into our ongoing program.

\end{abstract}

\maketitle  

\section{Introduction}
\label{sec:introduction}

The broad application of effective field theories like the chiral Lagrangian
to the description of low energy phenomena in QCD has been one of the major
theoretical advances of the last quarter century \cite{Gasser:1985gg}.
One substantial limitation of this approach is that the chiral Lagrangian
has additional symmetry \cite{Kaplan:1986ru} that precludes the determination
of all the low energy constants (LEC's) of the effective theory
using only experimental input.  This problem is resolved by computing
low energy observables directly in QCD and fixing the undetermined LEC's
by matching to predictions of chiral perturbation theory (ChPT).

Unfortunately, the low energy regime of QCD is non-perturbative,
so relevant observables are not so easily computed.  This has led
to an entire industry of estimating LEC's (with uncontrolled systematic errors)
using model calculations and phenomenology, culminating
in a generally accepted set of values \cite{Bijnens:1994qh}.  Lattice QCD
has also made some inroads in this area, with its own set
of systematic errors, including errors due to $N_f\ne 3$ and dynamical quarks
that are too heavy.  See Wittig's review talk for a current summary
\cite{Wittig:2002}.

Recently it was realized that ChPT and the corresponding
partially quenched chiral perturbation theory (pqChPT)
\cite{Bernard:1992mk,Bernard:1994sv} share the same low energy constants
\cite{Sharpe:2000bc,Cohen:1999kk}.  As partially quenched QCD (pqQCD)
simulations are somewhat less computationally demanding, recent efforts
have been made to determine the LEC's, particularly those affecting the mass
of the up quark \cite{Irving:2001vy}.  However, as the number of light
dynamical sea quarks $N_f$ doesn't appear explicitly in the chiral Lagrangian,
LEC's computed with $N_f \ne 3$ have no \textit{a priori} relation
to those of phenomenological interest, although \textit{a posteriori}
evidence suggests that the $N_f$ dependence may be mild.
Here, we will review our recent $N_f=3$ calculations \cite{Nelson:2001tb},
emphasizing our program to estimate all expected systematic errors, and discuss
our plans for reducing the systematic errors in the future.

\section{LEC's and systematic error}
\label{sec:LECs}

Our goal is to determine the low energy constants $2 \alpha_8 - \alpha_5$
and $2 \alpha_6 - \alpha_4$ \footnote{$\alpha_i \equiv 8(4\pi)^2 L_i$}
by studying the sea and valence quark mass ($m_S$ and $m_V$) dependence
of the pseudo-Goldstone meson mass $\widetilde{M}_\pi$ and decay constant
$f_\pi$.  NLO pqChPT predicts the dependence in terms of the LEC's, with loops
cut off at $\Lambda_\chi \equiv 4 \pi f$, as
\begin{eqnarray}
\widetilde{M}_\pi^2 & = & z \Lambda_\chi^2 m_V \left[
  1 + z m_V \left( 2 \alpha_8 - \alpha_5 + \frac{1}{N_f} \right)
\right. \nonumber \\*
& & + 2 m_S N_f \left( 2 \alpha_6 - \alpha_4 - \frac{1}{N_f^2} \right)
\nonumber \\*
& & + \left.
  \frac{z}{N_f} \left( 2 m_V - m_S \right) \log \left( m_V + m_S \right)
\right]
\end{eqnarray}
\begin{eqnarray}
f_\pi & = & f \left[
  1 + \frac{\alpha_5}{2} z m_V + \frac{\alpha_4}{2} z m_S N_f
\right. \nonumber \\*
& & + \left.
  \frac{z N_f}{4} \left( m_V + m_S \right)
  \log \frac{z}{2} \left( m_V + m_S \right)
\right]
\end{eqnarray}
where $z \equiv 2 \mu / \Lambda_\chi^2$ is related to the chiral condensate
$\mu$ and $\widetilde{M}$ refers to a QCD meson mass,
without QED contributions.  Similar formulae for fixed $m_S$ can be used
to extract only $\alpha_5$ and $2 \alpha_8 - \alpha_5$.  We refer readers
first to our earlier work \cite{Nelson:2001tb} and focus on details
not covered there.

The primary sources of systematic error are finite volume, finite lattice
spacing, finite step size, flavor (or ``taste'' \cite{Aubin:2002ss})
symmetry violations and the choice of sea and valence quark masses
for observables included in the fitting procedure.  The errors
are not independent.  For example, as smaller quark masses are included
in the analysis, larger volumes are required to minimize finite volume
errors.  Also, at smaller lattice spacings all systematic errors due
to lattice artifacts are reduced, including flavor symmetry violations.
So, while not justified by the preceding observation, we have chosen
to make generous estimates of our systematic errors and add them
in quadrature.

Our procedure for estimating systematic errors is simple.  Holding
all other parameters fixed, we independently vary the volume, step size,
lattice spacing and sea quark mass.  To estimate fitting systematics,
we vary the range of quark masses included in the fit, ensuring that all
masses are within the expected radius of convergence of the effective theory.
To estimate flavor symmetry breaking still remaining
at a given lattice spacing, we perform HYP blocking \cite{Hasenfratz:2001hp}
of the configurations and recompute the observables.  This procedure may seem
\textit{ad hoc} but it certainly has the correct continuum limit
as the blocking becomes irrelevant for sufficiently smooth gauge fields.

As our paper shows, this recipe for estimating systematic errors works well
with $\widetilde{M}_\pi$.  HYP blocking doesn't significantly
change the leading order behavior, as expected since local blocking
shouldn't effect quantities measured at long distances on the lattice.
In the presence of flavor breaking, the pions contributing to the chiral logs
are too heavy and systematically shift NLO terms away from their
continuum values.  With HYP blocking, these pions become more degenerate
and the systematic error is reduced.

\begin{figure}[ht]
\label{fig:hyp_16+16.f_pi}
\includegraphics[width=0.46\textwidth]{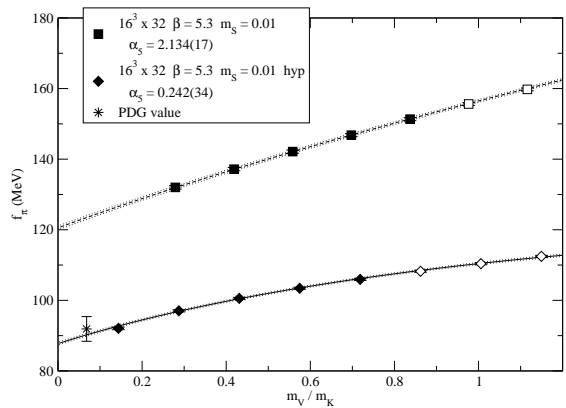}
\vspace{-4ex}
\caption{$f_\pi$ with and without HYP blocking on $16^3\times 32$,
  $\beta=5.3$, and $m_S=0.01$.}
\vspace{-4ex}
\end{figure}

An interesting contrast is what happens to $f_\pi$.  Since the dominant
contributions are local, Fig.~\ref{fig:hyp_16+16.f_pi} clearly shows
the effect of HYP blocking on the leading order behavior of $f_\pi$.
%
%
This seems disconcerting at first glance until one notices the unimproved
staggered $f_\pi$ is substantially higher than the PDG value whereas
the HYP blocked $f_\pi$ is consistent with it.  The real difficulty comes
when extracting $\alpha_5$: for unimproved staggered $\alpha_5 = 2.134(17)$
and for HYP $\alpha_5 = 0.242(34)$.  It doesn't seem reasonable to treat
the difference in the central values as an estimate of systematic error.
Luckily, $\alpha_5$ is not affected by the Kaplan--Manohar ambiguity
so phenomenological estimates are likely to be more reliable.  On the other
hand, these estimates currently have 20\% errors so we would like
to consider how to better determine LEC's like $\alpha_5$ in the next section.

\section{Future plans}

\begin{figure}[ht]
\label{fig:R_M_vs_m_v_8_X_full_hyp}
\includegraphics[width=0.46\textwidth]{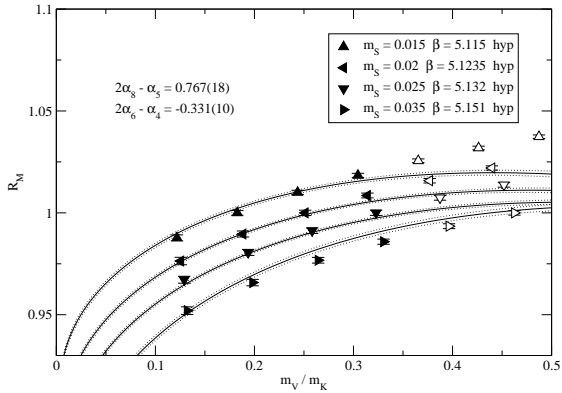}
\vspace{-4ex}
\caption{$R_M \equiv \widetilde{M}^2_\pi(m_S) m_V
  / \widetilde{M}^2_\pi(m_V) m_S$ on $8^3\times 32$ lattices
  at $a^{-1} \approx 0.7~\mathrm{GeV}$}
\vspace{-4ex}
\end{figure}

Our original calculation \cite{Nelson:2001tb} centered around a simulation
of a single dynamical sea quark mass at $a^{-1} \approx 1.3~\mathrm{GeV}$,
thus it was not possible to extract $2 \alpha_6 - \alpha_4$.  However,
our study included supporting simulations with several different sea quark
masses at $a^{-1} \approx 0.7~\mathrm{GeV}$ on comparable physical volumes.
So, we offer Fig.~\ref{fig:R_M_vs_m_v_8_X_full_hyp} as a demonstration
of the technique to extract $2 \alpha_6 - \alpha_4$ and $2 \alpha_8 - \alpha_5$
simultaneously.
%
%
This figure shows a single fit to four simulations
with $\chi^2/\mathrm{dof} \sim 1$ and should be compared to the corresponding
figure in \cite{Nelson:2001tb} where each simulation is fit independently.
We expect to perform a detailed extraction of $2 \alpha_6 - \alpha_4$
by this method, including a study of systematic errors, when sufficient
data becomes available at smaller lattice spacings.

In section \ref{sec:LECs}, we argued that HYP blocking is useful
for estimating systematic errors due to flavor symmetry breaking.
At this conference, we have heard about recent efforts
to find an efficient dynamical HYP algorithm
\cite{Alexandru:2002sw,Hasenfratz:2002pt}.  We hope to
generate an ensemble of $N_f = 3$ dynamical HYP lattices which will be used
to calculate kaon weak matrix elements for $\epsilon^\prime / \epsilon$
\cite{Bhattacharya:2002zw} as well as Gasser--Leutwyler coefficients.
This should help resolve such questions as how to estimate the systematics
of flavor breaking in quantities like $f_\pi$.

Also at this conference, Bernard provided an update on his program
to incorporate flavor symmetry breaking into ChPT \cite{Aubin:2002ss}.
This is an important development for our program.  It altogether eliminates
flavor breaking as a source of systematic error as the effects are
correctly modeled in our fitting function.  It may also increase the range
over which the data are well represented by our fitting functions
(still within the expected radius of convergence, of course).

Finally, we note that while our determination of $2 \alpha_8 - \alpha_5$
is completely dominated by systematic error, its systematic error
is not the dominant source of uncertainty in the ratio $m_u / m_d$.
The dominant uncertainty comes from the unknown electromagnetic contributions
to the masses of the pseudoscalar mesons.  A technique was devised
to measure these effects in quenched lattice simulations \cite{Duncan:1996xy}
and was presented at Lattice '96 \cite{Duncan:1997sq}.  We intend to explore
the extension of this work to simulations with dynamical fermions and hope
to present our first results at Lattice '03.

\bibliography{main}

\end{document}